\documentclass[12pt]{article}
\usepackage{color}

\newcommand{\beq}{\begin{equation}}
\newcommand{\eeq}{\end{equation}}
\newcommand{\beqa}{\begin{eqnarray}}
\newcommand{\eeqa}{\end{eqnarray}}
\newcommand{\beqar}{\begin{eqnarray*}}
\newcommand{\eeqar}{\end{eqnarray*}}

\newcommand{\al}{\alpha}
\newcommand{\be}{\beta}

\def\spa          {\ \ \ }
\def\non          {\nonumber}
\def\ha           {\mbox{$\frac{1}{2}$}}

\def\spa          {\ \ \ }
\def\mand         {\spa\mbox{and}\spa}

\def\Tr           {\mbox{\rm Tr}\,}
\def\STr          {\mbox{\rm STr}\,}
\def\Str          {\mbox{\rm Str}\,}

\def\cd           {{\cdot}}
\def\ran          {\rangle}
\def\lan          {\langle}

\def\fsH    {H\!\!\!\!/\,}
\newcommand{\del}{\delta}

\newcommand{\eps}{\epsilon}
\newcommand{\ga}{\gamma}
\newcommand{\Ga}{\Gamma}

\newcommand{\inn}{\!\cdot\!}

\newcommand{\lam}{\lambda}

\newcommand{\sig}{\sigma}
\newcommand\bPsi{{\bar \Psi }}

\newcommand{\z}{\zeta}

\newcommand{\ie}{{\it i.e.,}\ }
\newcommand{\labell}[1]{\label{#1}} 
\newcommand{\reef}[1]{(\ref{#1})}
\newcommand\prt{\partial}

\def\sst#1{{\scriptscriptstyle #1}}
\def\0{{\sst{(0)}}}
\def\1{{\sst{(1)}}}
\def\2{{\sst{(2)}}}
\def\3{{\sst{(3)}}}
\def\4{{\sst{(4)}}}
\def\5{{\sst{(5)}}}
\def\6{{\sst{(6)}}}
\def\7{{\sst{(7)}}}
\def\8{{\sst{(8)}}}

\begin{document}
\baselineskip 18pt%
\begin{titlepage}
\vspace*{1mm}%
\hfill
\vbox{

    \halign{#\hfil         \cr
           } 
      }  
\vspace*{9mm}
\vspace*{9mm}%

\center{{\bf\Large Selection Rules and  RR Couplings on Non-BPS Branes 
}}\vspace*{1mm} \centerline{{\Large {\bf  }}}
\begin{center}
{Ehsan Hatefi }

\vspace*{0.8cm}{ {\it
International Centre for Theoretical Physics\\
 Strada Costiera 11, Trieste, Italy  }}
\footnote{ehatefi@ictp.it}

\vspace*{0.1cm}
\vspace*{.1cm}
\end{center}
\begin{center}{\bf Abstract}\end{center}
\begin{quote}

We compute three and four point functions of the non-BPS scattering amplitudes, including a closed string Ramond-Ramond, gauge/scalar and tachyon in type IIA (IIB) superstring theories. We then discover a unique expansion for tachyon amplitudes in both non-BPS and D-brane anti D-brane formalisms. Based on remarks on Chan-Paton factors and arXiv:1304.3711, we propose selection rules for all non-BPS scattering amplitudes of type II superstring theory. These selection rules, rule out various non-BPS higher point  correlation functions of the string theory.

\end{quote}
\end{titlepage}

\section{Introduction}

 It is shown  that $D_{p}$-branes  have to be realised  as sources for Ramond-Ramond field (RR) for all kinds of either non-BPS  or stable BPS D-branes in string theory  \cite{Polchinski:1995mt,Witten:1995im}.
Through RR couplings several important subjects have been addressed.  Let us just introduce some of the most important ones. The phenomenon brane inside branes  \cite{Douglas:1995bn,Douglas:1997ch,Li:1995pq,Green:1996dd}, K-theory (of course with D-brane language) \cite{Witten:1998cd}, Dielectric effect (which is known to be Myers effect  \cite{Myers:1999ps}), all order higher derivative corrections to Myers effect for stable branes  \cite{Hatefi:2012zh,Hatefi:2012ve,Hatefi:2010ik} and more importantly all corrections to non-BPS branes in \cite{Hatefi:2013mwa} are found out.

\vskip .1in

To begin with non-BPS branes in IIA (IIB) theory, $D_{p}$-branes with  $p$ (spatial dimension of branes) odd (even) should be considered.  In \cite{Hatefi:2012rx} we obtained a conjecture for an infinite number of the higher derivative corrections to all orders in $\alpha'$ for both BPS and non-BPS amplitudes. In a very recent paper,  \cite{Hatefi:2013eia} investigated in detail this conjecture at some fundamental levels that can be applied to fermionic amplitudes as well.
 \vskip .1in

 In the case of non-BPS branes and their effective actions, it is known  that the spectrum of non-BPS branes includes all massless, some massive and tachyon strings \cite{Sen:2004nf}.  Specifically diverse discussions have clarified that the effective theory of unstable branes must have just massless and tachyon states as one needs to integrate out all massive modes. Their effective actions indeed have two parts involving DBI and Wess-Zumino part.

\vskip .1in

There are some methods to actually obtain these effective actions, such as Boundary String Field theory  (BSFT) and S-Matrix method. Apart from basic references for BSFT \cite{Kraus:2000nj,Takayanagi:2000rz},  BSFT  has already been explained in the introduction of \cite{Garousi:2008ge}.  On the other hand the Wess-Zumino  part is given in \cite{Hatefi:2012wj} as follows
\beqa
S_{WZ}&=&\mu_p' \int_{\Sigma_{(p+1)}} C \wedge \Str e^{i2\pi\alpha'\cal F}\labell{WZ'},\eeqa
with   super connection's curvature as
 \begin{displaymath}
i{\cal F} = \left(
\begin{array}{cc}
iF -\beta'^2 T^2 & \beta' DT \\
\beta' DT & iF -\beta'^2T^2
\end{array}
\right) \ ,
\non\end{displaymath}
where  $\beta'$  is a normalisation constant.  

The second method to begin with  non-BPS branes effective actions  is S-matrix method. This approach may be used to discover new couplings with exact $\alpha'$ corrections without any on-shell ambiguity. If one works with this method then one is able to fix all the coefficients of the higher derivative corrections to all orders in $\alpha'$.  Notice that the internal CP matrix of tachyons around unstable point of tachyon DBI has to be taken into account, otherwise the computations of the S-matrices in type II string theory do not make sense.  Basically open string tachyons  in  zero picture come with  $\sigma_1$ while in (-1) picture they carry  $\sigma_2$ Pauli matrix.
The correct form of tachyon DBI effective action including its internal  CP matrices is introduced in \cite{Garousi:2008ge,Hatefi:2012wj}
\beqa
S_{DBI}&\sim&\int
d^{p+1}\sigma \STr\left(\frac{}{}V({ T^iT^i})\sqrt{1+\frac{1}{2}[T^i,T^j][T^j,T^i])}\right.\labell{nonab} \\
&&\qquad\qquad\left.
\times\sqrt{-\det(\eta_{ab}
+2\pi\alpha'F_{ab}+2\pi\alpha'D_a{ T^i}(Q^{-1})^{ij}D_b{ T^j})} \right)\,,\nonumber\eeqa  with
\beqa
V({T^iT^i})&=&e^{-\pi{ T^iT^i}/2}, \quad Q^{ij}=I\delta^{ij}-i[T^i,T^j],\eeqa

where $i,j=1,2$, \ie $T^1=T\sigma_1$, $T^2=T\sigma_2$  (more information can be found in the next section). Since the total  super ghost charge must be (-2) for disk amplitudes, one  realises that by expanding the square roots, one has to keep two tachyons in (-1) pictures and they have  to be carried with $T^2$. Symmetric trace is also needed.

 Surprisingly, this action around tachyon potential's stable point , takes the usual tachyon DBI action \cite{Sen:1999md,Kluson:2000iy} with $T^4V(T^2)$ potential. Note that by sending tachyon  to infinity, the term  $T^4V(TT)$ will send  to zero. This is also expected from  condensing an unstable brane. It is worth mentioning that \reef{nonab}  makes consistent results with the computations of the amplitudes of one RR $(C)$, one tachyon $(T)$ and two scalar fields $(\phi \phi)$ including all their infinite higher derivative corrections  \cite{Hatefi:2012wj}.

\vskip.1in

On the other hand based on direct S-matrix computations of  $(CTTA)$ amplitude in the world volume of brane -anti brane system, the correct form of D-brane anti D-brane  $(D\bar D)$ effective action (symmetrized trace effective action) with all infinite higher derivative corrections of two tachyons and two  gauge fields is derived in \cite{Garousi:2007fk}. More recently by making use of the direct computations, all order $\alpha'$ higher derivative corrections to two tachyons and two scalars  in the world volume of brane -anti brane system are discovered in \cite{Hatefi:2012cp}.
To look for some of the applications of the higher derivative corrections for BPS branes in M-theory  \cite{Hatefi:2012sy} (considering the $N^3$ entropy growth of $M5$ branes and Myers effect) and \cite{McOrist:2012yc} are suggested. For some other applications to all order corrections  in gauge-gravity duality  \cite{Hatefi:2012bp} is proposed.
The importance of the higher derivative corrections  for tachyon amplitudes just around the unstable point of tachyon action is also explained in\cite{Hatefi:2012wj}.  Subsequently, using direct conformal field theory calculations tachyon DBI supersymmetrized action (\cite{Hatefi:2013mwa},\cite{Aganagic:1996nn,Aganagic:1997zk}) is obtained. 
\beqa
L=-T_pV(T)\sqrt{-\det(\eta_{ab}+2\pi\alpha'F_{ab}-2\pi\alpha'\bar\Psi\ga_b\prt_a\Psi+\pi^2\alpha'^2\bar\Psi\ga^{\mu}\prt_a\Psi\bar\Psi\ga_{\mu}\prt_b\Psi + 2\pi\alpha'\prt_aT\prt_bT )}
\nonumber
\eeqa

In  \cite{Hatefi:2013mwa}, we have explored  all order $\alpha'$ higher derivative corrections to two tachyons and two fermions as well.  Employing unstable branes  might help us to  realise some properties of the superstring theory in time-dependent backgrounds \cite{Gutperle:2002ai,Sen:2002in,Sen:2004nf,Lambert:2003zr}. It is known that tachyons  are the sources of the instabilities in flat space thus  by studying them in an effective field theory, we may be able to get some fascinating results. Indeed tachyon Born-Infeld effective action \cite{Kluson:2000iy} in string theory can explain some of the decay properties of non-BPS D$_p$-branes \cite{Sen:2002in,Sen:2002an}. 
\noindent 
If one works with this effective action then one may understand the evolution of these unstable branes in time-dependent backgrounds. One can also deal with the possible cosmological applications of the tachyonic DBI action.  In order to start with the explanation of the inflation in string theory \cite{Dvali:1998pa,Burgess:2001fx,Kachru:2003sx}, we need to know the effective action of $D\bar D$  system  involving its all order $\alpha'$ higher derivative corrections \cite{Garousi:2007fk,Hatefi:2012cp}. Some other applications on unstable branes are considered in \cite{Choudhury:2003vr}. To work with spontaneous chiral symmetry breaking (SCSB), tachyons and their higher derivative corrections are used in diverse models such as holographic model of the QCD \cite{Casero:2007ae,Dhar:2007bz,Dhar:2008um}. If we take the brane, anti brane system just as a background (to be dual to confined color theory) then one can introduce flavor branes \cite{Sakai:2004cn} where by decreasing the number of  flavor branes with respect to the number of color branes, the $D\bar D$  system may be  considered as a probe. 
 \vskip .1in

The outline of the  paper is as follows.
 \vskip .1in
 
In the next section  in the presence of non-BPS branes  and based on internal CP matrix of  all strings,  we are going to construct selection rules for non-BPS amplitudes. We observe that these selection rules must be used to rule out several non-BPS higher point correlation functions of  type IIA (IIB)  superstring theory without the need for  knowing world-sheet integrals. More significantly the proposed selection rules show that there should not be any couplings for some of the open/closed strings in the effective actions of non-BPS branes. In section three by carrying out explicit calculations  we obtain all order $\alpha'$ higher derivative corrections to the amplitude of one RR, one tachyon and one scalar field $<V_C V_T V_{\phi}>$ in the world volume of non-BPS branes.
 \vskip .1in

In particular we discover a unique expansion for tachyon amplitudes. This expansion is very useful because by applying it to string amplitudes, one finds out all the singularities of the non-BPS  higher point correlation functions of the string theory without  knowing the exact results of the world-sheet integrals.
This idea clearly had been applied to the amplitude of one RR and four tachyons in the world volume of $D\bar D$ system but has not been announced yet \cite{Hatefi:2013eh}. Using selection rules and universal tachyon expansion in section four, we will deal with the amplitude of one RR, one scalar, one gauge field and one tachyon in the world volume of non-BPS branes. 

Having used some of new Wess-Zumino terms, tachyonic DBI action, this universal tachyon expansion, the proposed selection rules and the derived  corrections of one RR, one tachyon and one scalar field (results in section four), we are able to exactly produce infinite number of $(u'=u+1/4)$- channel tachyon and $t$- channel massless scalar poles of $<V_C V_T V _A V_{\phi}>$ amplitude  as well as their higher derivative corrections. If we apply the selection rules to  the field theory of  $<V_C V_T V _A V_{\phi}>$ then we understand neither there should be  single / double $s',(s'+u'+t)$ channel  tachyon, gauge/ scalar poles nor infinite poles. Indeed this  $<V_C V_T V _A V_{\phi}>$ amplitude  is an exceptional S-matrix.
\section{Selection rules for non-BPS amplitudes}

Based on Chan-Paton factors and new proposal appeared in \cite{Hatefi:2013mwa}, we are going to propose the complete selection rules for non-BPS scattering amplitudes of type II superstring theory.  Without any knowledge of mixture of the closed-open string theory correlators, these rules can be used to show that some of the non-BPS higher point correlation functions of the string theory will be ruled out. 
\vskip .1in

As mentioned by \cite{Sen:1999mg}, one can read off the internal Chan-Paton  matrix of the   non-BPS D-brane open strings  from  D-brane anti D-brane Chan-Paton  matrix.  The open strings of the brane anti brane system (two real tachyons, massless scalar/gauge fields) are given by the following CP matrices
\beqa
(a):\,\pmatrix{0&0\cr
0&1},\,\,\,
(b):\,\pmatrix{1&0\cr
0&0},\,\,\, (c):\,\pmatrix{0&0\cr
1&0},\,\,\, (d):\,\pmatrix{0&1\cr
0&0}\labell{M121}, \eeqa
\noindent 
(a),(b) matrices are related to scalar/gauge located either on brane or anti brane and (c), (d) matrices show tachyons stretched from brane to anti brane and vice versa. The projection operator $(-1)^{F_L}$ does not play any role on the world-sheet fields; however, it changes  brane to anti brane. Hence the projection operator plays the role on  CP matrix $\Lambda$  as follows

\beqa
\Lambda\rightarrow \sigma_1\Lambda (\sigma_1)^{-1}, \nonumber\eeqa
Once the projection operator is applied, only the states carrying either  $I$  or $\sigma_1$ CP matrices will remain. Therefore it is easy to show that  all massless strings should carry   internal  $I$ CP matrix. In order to get consistent results with the non-BPS S-matrix elements in string theory, tachyon vertex operator  in zero picture must carry  $\sigma_1$ CP matrix \cite{Hatefi:2012wj}. Let us devote  $I$  CP matrix  to massless strings and $\sigma_1$ CP matrix to tachyon  vertex operator in the zero picture.  Since the picture changing operator for a non-BPS brane  carries $\sigma_3$ CP factor \cite{DeSmet:2000je}, one can apply it  to  all vertex operators in zero picture to obtain the CP factor of the strings in (-1)- picture.

Applying the picture changing operator on massless vertex operators in zero picture, we show that their internal CP matrix in (-1) picture is  $I\sigma_3=\sigma_3$ and particularly the CP matrix of tachyon in (-1) picture becomes  $\sigma_1\sigma_3\sim\sigma_2$. The amplitude of two fermions and one gauge field makes sense even in the presence of non-BPS branes. Given this fact, we  are able to fix the CP matrices of the fermion fields      
in  \cite{Hatefi:2013mwa} as follows  
\beqa
A^{\bPsi^{-1/2}\Psi^{-1/2} A^{-1}} &\sim& \Tr(\sigma_3 I \sigma_3) \neq 0.\nonumber\eeqa

\vskip .1in

Ramond-Ramond vertex operator carries CP matrix like other strings. Since there is a non zero coupling between two tachyons and one RR in the world volume of brane anti brane systems \cite{Kennedy:1999nn,Garousi:2007fk}, it has been pointed out in    \cite{Hatefi:2012wj} that the RR for brane anti brane systems in  $(-1/2,-1/2)$ picture  has to carry   $\sigma_3$  CP matrix.  It is also discussed in detail in \cite{Hatefi:2012wj} that due to  \cite{Sen:1999mg} there is an extra coefficient in RR vertex operator in (-1) picture for the non-BPS branes which means that RR in (-1) picture carries $\sigma_3\sigma_1$ for the non-BPS brane systems. Its CP factor in (-2) picture has also been addressed in \cite{Hatefi:2012wj}. As the closed strings (NS-NS sector) in $(0,0)$ picture come with $I$ internal CP matrix the internal CP factors can be verified  by the S-matrix element of one RR, one graviton and one tachyon. The field theory of non-BPS D-branes  reveals that the amplitude must have tachyon pole.
\vskip .1in

In field theory, this pole is found by the Feynman amplitude for which tachyon propagates between $CT$ and $TTh$ where $C$ and $h$ denote RR and graviton appropriately; however, there is no such pole in the field theory of brane anti brane  system (\ie there is no $CT$ coupling  in ($D\bar{D}$) case). Therefore, the internal CP factor  ($D\bar{D}$) system is zero but for the non-BPS case, the internal CP factor  is non-zero.  It is easy to see that for the RR vertex operator in $(-1/2,-1/2)$ along with graviton in $(0,0)$ and tachyon in $(-1)$ picture, the CP factor of the $CTh$ amplitude for $D\bar{D}$ case is $\Tr(\sigma_3I\sigma_2)=0$ while  its CP matrix is $\Tr(\sigma_3\sigma_1I\sigma_2)\neq 0$ for the non-BPS brane. Accordingly, when we choose the RR in $(-1/2,-1/2)$, graviton in $(-1,0)$ and tachyon in $0$ picture, then the CP factor for $D\bar{D}$ case is $\Tr(\sigma_3\sigma_3\sigma_1)=0$ while   its CP matrix  is  non-zero $(\Tr(\sigma_3\sigma_1\sigma_3\sigma_1)\neq 0$) for non-BPS brane. We are going to generalise these selection rules. 

\vskip .2in

Based on internal CP matrices  discussed in \cite{Hatefi:2013mwa}, there is no coupling between two closed string Ramond-Ramond fields and one tachyon in the world volume of non-BPS branes, because
\beqa
A^{C^{-1}C^{-1} T^{0}} &\sim& \Tr(\sigma_3\sigma_1\sigma_3\sigma_1 \sigma_1) =0 \label{jjk},\eeqa
Note that the correlation function of four spin operators and one fermion field in 10 dimensions is non zero;however, based on internal CP matrices in \reef{jjk}, there are no couplings between two RR's and one tachyon in type IIA(B) superstring theory.

\vskip .1in

 Now we can generalise these selection rules further. Since we know that the total super ghost charge must be (-2) for disk level S-matrix and scalars, gauges in zero picture carry (I) CP matrix, we conclude that the amplitude of two RR's, an arbitrary number of gauges/scalars and odd-number  of tachyons in type IIA(B) superstring theory is zero. Of course one can add some arbitrary number  of (NS-NS) to these amplitudes as well for which the result  remains zero again, \ie
\beqa
A^{C^{-1}C^{-1} A^0 \phi^0 \cdots A^0 T^{0} T^{0}\cdots T^{0}} &=& 0. \nonumber\eeqa
In addition to the above equation, we can see that many amplitudes have no contributions to superstring theory.
The amplitudes of one RR, some arbitrary number of   (NS-NS) closed string states and odd number of tachyons in the world volume of brane anti brane vanish. More significantly the CP factor of a RR, some arbitrary closed string   (NS-NS) states and even number of tachyons in the world volume of non-BPS branes is also zero, thus there are no such couplings in type IIA(B) superstring theories.

\vskip .2in

Based on internal CP matrices of the open strings in  \cite{Hatefi:2013mwa}, we show that the tree level S-matrix elements of two fermion fields and one tachyon in the world volume of non-BPS branes becomes zero result; however, the correlation function of this $(\bar\psi \psi T)$ amplitude (coming from two spin operators) , \ie
\beqa
<:S_{A}(x_1):S_{B}(x_2):>= x_{12}^{-5/4} (C^{-1})_{AB}\nonumber\eeqa
 is non zero. We can easily generalise and postulate these internal CP matrices to all the vertex operators as below:
\beqa
V_{T}^{(0)}(x) &=&  \alpha' ik\cd\psi(x) e^{\alpha' ik\cd X(x)}\lam\otimes\sigma_1,
\nonumber\\
V_{T}^{(-1)}(x) &=&e^{-\phi(x)} e^{\alpha' ik\cd X(x)}\lam\otimes\sigma_2\nonumber\\
V_\phi^{(-1)}(x)&=&e^{-\phi(x)}\xi_i\psi^i(x)e^{ \alpha'iq\inn X(x)}\lam\otimes \sigma_3 \nonumber\\
V_A^{(-1)}(x)&=&e^{-\phi(x)}\xi_a\psi^a(x)e^{ \alpha'iq\inn X(x)}\lam\otimes \sigma_3 \nonumber\\
V_{\phi}^{(0)}(x) &=& \xi_{1i}(\partial^i X(x)+i\alpha'k.\psi\psi^i(x))e^{\alpha'ik.X(x)}\otimes I\nonumber\\
V_{A}^{(0)}(x) &=& \xi_{1a}(\partial^a X(x)+i\alpha'k.\psi\psi^a(x))e^{\alpha'ik.X(x)}\otimes I\label{d4Vs}\\
V_{\bar\Psi}^{(-1/2)}(x)&=&\bar u^Ae^{-\phi(x)/2}S_A(x)\,e^{ \alpha'iq.X(x)}\lam\otimes\sigma_3 \nonumber\\
V_{\Psi}^{(-1/2)}(x)&=&u^Be^{-\phi(x)/2}S_B(x)\,e^{ \alpha'  iq.X(x)}\lam\otimes I \nonumber\\
V_{C}^{(-\frac{1}{2},-\frac{1}{2})}(z,\bar{z})&=&(P_{-}\fsH_{(n)}M_p)^{\alpha\beta}e^{-\phi(z)/2}
S_{\al}(z)e^{i\frac{\alpha'}{2}p\cd X(z)}e^{-\phi(\bar{z})/2} S_{\be}(\bar{z})
e^{i\frac{\alpha'}{2}p\cd D \cd X(\bar{z})}\lam\otimes\sigma_3\sigma_1,\nonumber
\eeqa

where the CP factor of RR  for non-BPS branes is $\sigma_3\sigma_1$, meanwhile its CP factor  for $D\bar D$ system is $\sigma_3$. Given  \cite{Hatefi:2013mwa}, we are going to generalise non-BPS fermionic amplitudes further.

\vskip .1in

The amplitude of  two fermions, an arbitrary number of scalar/gauge fields and odd number of tachyons in the world volume of non-BPS branes in superstring theory is given by the following correlation function : 

\beqa
{\cal A}^{\bPsi^{-1/2},\Psi^{-1/2},T^{-1},T^{0},\cdots, T^{0},A_{1}^0, \phi_{l}^0,\cdots A_{m}^0,\phi_{l}^0} & \sim & \int dx_1\cdots dx_{(2n+m+l+3)}
 \Tr\lan
V_{\bPsi}^{(-1/2)}(x_1)V_{\Psi}^{(-1/2)}(x_2)   \nonumber\\&&\times    V_{T}^{(-1)}(x_3)
V_T^{0}(x_4)
\cdots V_T^{0}(x_{2n+1})V_A^{0}(x_{m_1})V_{\phi}^{0}(x_{l_1})\nonumber\\&&\times
\cdots V_A^{0}(x_{m_m})V_{\phi}^{0}(x_{l_l})\ran,\nonumber\eeqa

We have shown in \cite{Hatefi:2012wj} that the correlator of two spin operators and some arbitrary number of fermion and/or currents in ten dimensions is non-zero. Thus one may guess that the above S-matrix element is non zero, but the internal CP factor of the above S-matrix is zero for all orderings of  tachyons, gauges and scalars, \ie
\beqa
{\cal A}^{\bPsi^{-1/2},\Psi^{-1/2},T^{-1},T^{0},\cdots, T^{0},A_{1}^0, \phi_{l}^0,\cdots A_{m}^0,\phi_{l}^0}&=& 0.\labell{zero2}
\eeqa
Hence  we discovered that in the world volume of non-BPS branes  (in both type IIA and IIB superstring theories) there are no couplings between  odd number of tachyons, two fermion fields and an arbitrary number of scalar/gauge vertex operators.

\vskip .1in

It is worth emphasising that these selection rules are very important to make sense of non-zero higher point correlation functions. Because by making use of them, we could remove several higher non-BPS correlators of the string theory without any knowledge of the world sheet integrals of the S-matrices. Now let us just talk about some other non-BPS three and four point functions of $C-\phi-T$ and $C-\phi-A-T$  amplitudes which make sense just in the world volume of non-BPS branes.

\section{The $T-\phi-C$ amplitude}

 In this section we are going to apply direct conformal field theory techniques \cite{Friedan:1985ge} to obtain  the three point amplitude between one RR, one  tachyon and one scalar field in the world volume of non-BPS branes. The amplitude is given by the following correlation function 
\beqa
{\cal A}^{T,\phi,RR} & \sim & \int dxdyd^2z
 \lan
V_{T}^{(0)}(y)V_\phi^{(-1)}(x)
V_{RR}^{(-1)}(z,\bar{z})\ran,\labell{cor1}\eeqa
where  tachyon,
 scalar field and RR vertex operators (including their CP factors) are  given in \reef{d4Vs}
such that all open strings have to be located in the boundary of the disk and RR must be replaced in the middle of the disk.

\vskip .2in

 $q,p,k_1$ are the momenta
of  scalar, RR and tachyon field which satisfy the following on-shell condition
\beqa
  q^2=p^2=0, \quad  k_{1}^2=1/4  ,\quad k_1.\xi_1=q.\xi_1=0,
\nonumber\eeqa
where the definitions of projector operator and RR field strength  are
\begin{displaymath}
P_{-} =\ha (1-\ga^{11}), \quad
\fsH_{(n)} = \frac{a
_n}{n!}H_{\mu_{1}\ldots\mu_{n}}\ga^{\mu_{1}}\ldots
\ga^{\mu_{n}},
\non\end{displaymath}

In type IIA  (type IIB) $n=2,4$,$a_n=i$  ($n=1,3,5$,$a_n=1$)  and also  the notation for the spinor is 
\beqa
(P_{-}\fsH_{(n)})^{\al\be} =
C^{\al\del}(P_{-}\fsH_{(n)})_{\del}{}^{\be}.
\nonumber
\eeqa

In order to make use of the holomorphic components of the world-sheet fields, one needs to work  with doubling trick. Thus we need to apply  the following change of variables 
\begin{displaymath}
\tilde{X}^{\mu}(\bar{z}) \rightarrow D^{\mu}_{\nu}X^{\nu}(\bar{z}) \ ,
\spa
\tilde{\psi}^{\mu}(\bar{z}) \rightarrow
D^{\mu}_{\nu}\psi^{\nu}(\bar{z}) \ ,
\spa
\tilde{\phi}(\bar{z}) \rightarrow \phi(\bar{z})\,, \mand
\tilde{S}_{\al}(\bar{z}) \rightarrow M_{\al}{}^{\be}{S}_{\be}(\bar{z})
 \ ,
\non\end{displaymath}

with the following matrices
\begin{displaymath}
D = \left( \begin{array}{cc}
-1_{9-p} & 0 \\
0 & 1_{p+1}
\end{array}
\right) \ ,\,\, \mand
M_p = \left\{\begin{array}{cc}\frac{\pm i}{(p+1)!}\ga^{i_{1}}\ga^{i_{2}}\ldots \ga^{i_{p+1}}
\eps_{i_{1}\ldots i_{p+1}}\,\,\,\,{\rm for\, p \,even}\\ \frac{\pm 1}{(p+1)!}\ga^{i_{1}}\ga^{i_{2}}\ldots \ga^{i_{p+1}}\ga_{11}
\eps_{i_{1}\ldots i_{p+1}} \,\,\,\,{\rm for\, p \,odd}\end{array}\right.
\non\end{displaymath}
\vskip .2in
Now we are allowed  to use just holomorphic part of the propagators for  $X^{\mu},\psi^\mu, \phi$, as follows
\begin{eqnarray}
\lan X^{\mu}(z)X^{\nu}(w)\ran & = & -\frac{\alpha'}{2}\eta^{\mu\nu}\log(z-w) \ , \non \\
\lan \psi^{\mu}(z)\psi^{\nu}(w) \ran & = & -\frac{\alpha'}{2}\eta^{\mu\nu}(z-w)^{-1} \ ,\non \\
\lan\phi(z)\phi(w)\ran & = & -\log(z-w) \ .
\labell{prop2}\end{eqnarray}

The $\sigma$-factor of the above S-matrix element is  $\Tr(\sigma_3\sigma_1\sigma_1\sigma_3)=2$ and the CP factor is $2\Tr(\lam_1\lam_2)$. If one considers the related vertices then  the amplitude is given as follows
\beqa
&&\int dx_1 dx_2 dx_4 dx_5   \alpha' (ik_{1a} \xi_{i}) (x_{24}x_{25})^{-1/2}(x_{45})^{-1/4} I_1
(P_{-}\fsH_{(n)}M_p)^{\alpha\beta} \nonumber\\&&\times
<:S_{\al}(x_4): S_{\be}(x_5):\psi^{a}(x_1):\psi^{i}(x_2):>,
 \nonumber\eeqa
where $x_4=z=x+iy,x_5=\bar z=x-iy$

\beqa
I_1&=&|x_{12}|^{\alpha'^2k_1.k_2}|x_{14}x_{15}|^{\frac{\alpha'^2}{2}k_1.p} |x_{24}x_{25}|^{ \frac{\alpha'^2}{2}  k_2.p}|x_{45}|^{\frac{\alpha'^2}{4}p.D.p},\nonumber\\
\eeqa

  Now using Wick-like rule one can derive the following correlator
  \beqa
  <:S_{\al}(x_4): S_{\be}(x_5):\psi^{a}(x_1):\psi^{i}(x_2):>&=& 2^{-1}(x_{14}x_{15}x_{24}x_{25})^{-1/2}(x_{45})^{-1/4}(\Gamma^{ia} C^{-1})_{\alpha\beta}.
  \nonumber\eeqa

 By applying this correlator into the above amplitude, we can easily show that the integrand is $SL(2,R)$ invariant.
 Let us do gauge fixing as  $(x_1,x_2,z,\bar z)=(x,-x,i,-i)$ so that now the integrand is proportional to\footnote{From now on we set $\alpha'=2$. }
\beqa
4k_{1a} \xi_i\int_{-\infty}^{\infty} dx (2x)^{-2u-1/2}
\bigg((1+x^{2})\bigg)^{-1/2 +2 u} \bigg(\Tr
(P_{-}\fsH_{(n)}M_p\Gamma^{ia})\bigg),\nonumber\eeqa

  where $u = -\frac{\alpha'}{2}(k_1+k_2)^2$ and particularly one has to apply the  conservation of momentum along the world volume of brane  ( $k_1^{a} + k_2^{a} + p^{a} =0$).

\vskip .2in

     The integral of the amplitude is
\beqa
{\cal A}^{T,\phi,RR} &=&(\pi\beta'\mu_p')2\sqrt{\pi} \frac{\Ga[-u+1/4]}{\Ga[3/4-u]}
 \Tr
(P_{-}\fsH_{(n)}M_p\Gamma^{ai})k_{1a}\xi_i\Tr(\lam_1\lam_2) \labell{amp383}\ .
\eeqa

The amplitude is normalised   by the coefficient of  $ (\pi\beta'\mu_p'/2)$ where  $ \beta' $  and   $  \mu_p' $ are Wess-Zumino normalisation and Ramond-Ramond charge of branes appropriately. The trace is non zero for $p+1= n$ and it is extracted as follows  
\beqa
\Tr\bigg(\fsH_{(n)}M_p
(\xi.\ga)(k_1.\ga)\bigg)\delta_{p+1,n}&=&\pm\frac{32}{(p+1)!}\eps^{a_{0}\cdots a_{p-1}a}H_{ia_{0}\cdots a_{p-1}}k_{1a}\xi_i
  \delta_{p+1,n}\nonumber\eeqa
  Given the fact that we want to compare S-matrix elements of $CT\phi$ with its field theory, we  do not fix the over all coefficient of the amplitude. Note that the trace involving $\gamma^{11}$ shows that the above results can be held for the following 
\beqa
  p>3 , H_n=*H_{10-n} , n\geq 5.
  \nonumber\eeqa

 It has been discussed in detail in \cite{Hatefi:2012wj} that the momentum expansion of any amplitude including tachyons should be obtained either with $k_i.k_j\rightarrow 0$ or $(k_i+k_j)^2\rightarrow 0$ where the latter appears just for massless channel pole. In the other words momentum expansion can be read by analysing  massless or tachyon poles of the amplitude. Nevertheless in this amplitude for both $u \rightarrow 0$ and $u \rightarrow -1/4$ (mass of tachyon), we observe that there are no massless/tachyon  poles in the  Gamma functions inside the amplitude of  \reef{amp383}.  Thus we might wonder about the momentum expansion of this amplitude.
 It is argued  in \cite{Hatefi:2012wj} that the  momentum expansion of non- BPS branes makes sense just in the presence of the following constraint:
\beqa
u=-p^ap_a\rightarrow \frac{-1}{4}.
\eeqa
while for  D-brane anti D-brane  system  $p^a p_a\rightarrow 0$ makes sense in the S-matrix elements of type II superstring theory  \cite{Hatefi:2012cp}.
\vskip .2in

 This $C\phi T$ amplitude makes sense just  for non-BPS branes , thus if we apply the momentum conservation along the world volume of brane then  one understands that the correct expansion is $u\rightarrow -1/4$, or in terms of momenta the expansion is around $k_1\inn k_2\rightarrow 0$. 
 
 Having expanded the Gamma functions that appeared in the amplitude we obtain an infinite number of higher derivative couplings of one scalar, one tachyon and one $C_{p}$ field. They are related to the  higher derivative extensions of the following Wess-Zumino  coupling:
  \beqa
    2i\beta'\mu'_p (2\pi\alpha')^2\int_{\Sigma_{p+1}} \bigg(\Tr(\partial_{i}C_{p}\wedge DT\phi^i)\bigg),
    \label{jj}\eeqa

In the above coupling Taylor expansion for scalar field has been employed.  The infinite higher derivative corrections of the above coupling must be obtained by considering  the momentum expansion inside  the S-matrix. The expansion is
\beqa
\sqrt{\pi}\frac{\Ga[-u+1/4]}{\Ga[3/4-u]}
 &=&\pi \sum_{m=-1}^{\infty}c_m(u+1/4)^{m+1}
\ .\labell{taylor61}\nonumber
\eeqa
 with the following coefficients
\beqa
c_{-1}&=&1,\nonumber\\
c_0&=&2ln(2),\nonumber\\
c_1&=&\frac{\pi}{6}(\pi^2+12ln(2)^2),\nonumber\\
c_2&=&\frac{1}{3}(6\z(3)+\pi^2ln(2)+4ln(2)^3), \nonumber\\
c_3&=&\frac{1}{360}(1440\z(3)ln(2)+120\pi^2 ln(2)^2+19\pi^4+240 ln(2)^4).\eeqa

  Note that the  above coefficients  are different from  the coefficients that appeared in the momentum expansion of the S-matrix element of one RR and two gauge fields (one tachyon and one gauge field) . If we replace the derived expansion in the final result of the amplitude and compare it with field theory coupling \reef{jj} then  we realise that the term including $c_{-1}$ can be precisely produced by \reef{jj}. On the other hand the higher derivative corrections of \reef{jj}  can be read with similar way as  follows
\beqa
\frac{2\beta'\mu_p'}{p!}(2\pi\alpha')^2\epsilon^{a_{0}...a_{p}} \partial_{i}C_{a_{0}...a_{p-1}}\wedge \Tr\left(\sum_{m=-1}^{\infty}c_m(\alpha')^{m+1}  D_{a_1}\cdots D_{a_{m+1}}D_{a_{p}}T D^{a_1}...D^{a_{m+1}}\phi^i\right) \labell{highaa}\eeqa
where in the above coupling all commutator terms should be ignored because we are looking for the infinite couplings of one RR, one scalar and one tachyon in the world volume of non-BPS branes.

\section{The $C-\phi-A-T$ amplitude}

\subsection{The $C^{-1}\phi^{-1}A^{0}T^{0}$ amplitude}

 In this section we are going to look for the closed form of the amplitude  of one RR, one scalar, one gauge and one  tachyon in the world volume of non-BPS branes in type IIA (IIB) superstring theory. This  $<V_C V_{\phi} V_A V_T>$ amplitude is a four  point (technically five point) function. It is pointed out in \cite{Hatefi:2012wj} that the vertex operators for non-BPS D-branes carry internal Chan-Paton factor. We know that  the S-matrix element of  BPS states does not depend on their pictures but  for non BPS branes the amplitude is just independent of  the picture of the vertex operators  if and only if we take into account the CP matrices inside the vertex operators.  Thus  $<V_C V_{\phi} V_A V_T>$  is  given by the following correlation function :

\beqa
{\cal A}^{C\phi AT} & \sim & \int dx_{1}dx_{2}dx_{3}dzd\bar{z}\,
  \lan V_{\phi}^{(-1)}{(x_{1})}
V_{A}^{(0)}{(x_{2})}V_T^{(0)}{(x_{3})}
V_{RR}^{(-\frac{1}{2},-\frac{1}{2})}(z,\bar{z})\ran,\labell{cor10}\eeqa
such that all open strings have to be located in the boundary of the disk and RR must be replaced in the middle of the disk. $k_1,k_2,p,k_3$ are the momenta
of scalar, gauge, RR and tachyon field which satisfy the following on-shell condition
\beqa
 k_{1}^2=k_{2}^2=p^2=0, \quad  k_{3}^2=1/4  ,\quad  k_2.\xi_2=k_2.\xi_1=k_1.\xi_1=k_3.\xi_1=0
\nonumber\eeqa

Obviously the definitions of the projector, the field strength of RR, D and $M_p$ matrices
are exactly  the same definitions that appeared in the last section. We also use holomorphic correlators \reef{prop2}.
\vskip .1in

If we consider the vertex operators then the  amplitude  is given by

\beqa {\cal A}^{C\phi A T}&\sim& \int
 dx_{1}dx_{2}dx_{3}dx_{4} dx_{5}\,
(P_{-}\fsH_{(n)}M_p)^{\al\be}\xi_{1i}\xi_{2a}(\alpha'ik_{3c})x_{45}^{-1/4}(x_{14}x_{15})^{-1/2}\nonumber\\&&
\times(I_1+I_2)\Tr(\lam_1\lam_2\lam_3)\Tr(\sig_3\sig_1\sig_3I\sig_1),\labell{1255}\eeqa

 where $x_{ij}=x_i-x_j$ and by applying the Wick theorem one finds the correlators as follows

\beqa
I_1&=&{<:e^{\alpha'ik_1.X(x_1)}: \partial X^a{(x_2)}e^{\alpha'ik_2.X(x_2)}
 :e^{\alpha'ik_3.X(x_3)}:e^{\frac{\alpha'}{2}ip.X(x_4)}:e^{\frac{\alpha'}{2}ip.D.X(x_5)}:>}\nonumber \\&&\times{<:S_{\al}(x_4):S_{\be}(x_5):\psi^i(x_1):\psi^c(x_3)>},\nonumber\\
I_2&=&{<:e^{\alpha'ik_1.X(x_1)}: e^{\alpha'ik_2.X(x_2)}
 :e^{\alpha'ik_3.X(x_3)}:e^{\frac{\alpha'}{2}ip.X(x_4)}:e^{\frac{\alpha'}{2}ip.D.X(x_5)}:>}\nonumber \\&&
\alpha'ik_{2d} {<:S_{\al}(x_4):S_{\be}(x_5):\psi^i(x_1):\psi^d\psi^{a}(x_2):\psi^c(x_3):>},
\eeqa

We need to  take into account  the Wick-theorem and \reef{prop2} to be able to compute the correlators of $X$.  In order to obtain the correlation function  including  two spin operators, one current and two fermion fields we use Wick-like rule~\cite{Liu:2001qa,Kostelecky:1986xg} and its generalisation  ~\cite{Hatefi:2010ik}.
First of all we need to find the following correlator 
\beqa
I_3^{ci}&=&<:S_{\al}(x_4):S_{\be}(x_5):\psi^i(x_1):\psi^c(x_3):>
=2^{-1}x_{45}^{-1/4} (x_{14}x_{15}x_{34}x_{35})^{-1/2}\bigg\{(\Gamma^{ci}C^{-1})_{\alpha\beta}\bigg\}.
\nonumber
\label{68}\eeqa

The computation of the correlation function of a current, two fermions  and two spin operators is really tedious, however, concerning the generalisation of Wick-Like rule ~\cite{Hatefi:2010ik} now one can easily obtain it as below

\beqa
I_4^{cadi}&=&<:S_{\al}(x_4):S_{\be}(x_5):\psi^i(x_1):\psi^d\psi^a(x_2):\psi^c(x_3):>\nonumber\\&&
=\bigg\{(\Gamma^{cadi}C^{-1})_{{\alpha\beta}}+\frac{Re[x_{24}x_{35}]}{x_{23}x_{45}}(2\eta^{dc}(\Gamma^{ai}C^{-1})_{{\alpha\beta}}-2\eta^{ac}(\Gamma^{di}C^{-1})_{{\alpha\beta}})\bigg\}\nonumber\\&&
2^{-2}x_{45}^{3/4}(x_{14}x_{15}x_{34}x_{35})^{-1/2}(x_{24}x_{25})^{-1}.\label{hh}\eeqa

We substitute all the related spin correlators in \reef{1255} and derive the general form of the amplitude as follows
\beqa
{\cal A}^{C\phi AT}&\!\!\!\!\sim\!\!\!\!\!&\int dx_{1}dx_{2} dx_{3}dx_{4}dx_{5}(P_{-}\fsH_{(n)}M_p)^{\al\be}I\xi_{1i}\xi_{2a}(-2\alpha'ik_{3c}) x_{45}^{-1/4}(x_{14}x_{15})^{-1/2}\nonumber\\&&\times
\bigg(a^a_1 I_3^{ci} +\alpha'ik_{2d}I_4^{cadi}\bigg)\Tr(\lam_1\lam_2\lam_3)\labell{amp3},\eeqa
such that
\beqa
I&=&|x_{12}|^{\alpha'^2k_1.k_2}|x_{13}|^{\alpha'^2k_1.k_3}|x_{14}x_{15}|^{\frac{\alpha'^2}{2}k_1.p}|x_{23}|^{\alpha'^2k_2.k_3}|x_{24}x_{25}|^{ \frac{\alpha'^2}{2}  k_2.p}
|x_{34}x_{35}|^{\frac{\alpha'^2}{2}   k_3.p}|x_{45}|^{\frac{\alpha'^2}{4}    p.D.p},\nonumber\\
a^a_1&=&ik_1^{a}\bigg(\frac{x_{14}}{x_{12}x_{24}}+\frac{x_{15}}{x_{12}x_{25}}\bigg)
+ik_3^{a}\bigg(\frac{x_{43}}{x_{23}x_{24}}+\frac{x_{53}}{x_{23}x_{25}}\bigg).
\eeqa

Now the amplitude is written such that it is explicitly SL(2,R) invariant. By gauge fixing  the location of open strings as
 \beqa
 x_{1}=0 ,\qquad x_{2}=1,\qquad x_{3}\rightarrow \infty,
 \eeqa
 we reach to the following integrals on the upper half plane

\beqa
 \int d^2 \!z |1-z|^{a} |z|^{b} (z - \bar{z})^{c}
(z + \bar{z})^{d},
 \eeqa

 where  $a,b,c$ are related to the Mandelstam variables as below
\beqar
s&=&-\frac{\alpha'}{2}(k_1+k_3)^2,\qquad t=-\frac{\alpha'}{2}(k_1+k_2)^2,\qquad u=-\frac{\alpha'}{2}(k_2+k_3)^2.
\qquad\eeqar
where for $d=0,1$ and for $d=2$ the results are derived accordingly in  \cite{Fotopoulos:2001pt}, \cite{Garousi:2008ge}. Given the solutions for integrals \cite{Fotopoulos:2001pt,Garousi:2008ge}, one can write the final result of the amplitude
\reef{1255} as
\beqa {\cal A}^{C\phi A T}&=&{\cal A}_{1}+{\cal A}_{2}+{\cal A}_{3},\labell{15u}\eeqa
where
\beqa
{\cal A}_{1}&\!\!\!\sim\!\!\!&2\xi_{1i}\xi_{2a}k_{3c}k_{2d}
\Tr(P_{-}\fsH_{(n)}M_p\Gamma^{cadi}
)L_1,
\nonumber\\
{\cal A}_{2}&\sim& 2\bigg\{\Tr(P_{-}\fsH_{(n)}M_p \gamma.\xi_2\gamma.\xi_{1})(u+\frac{1}{4})+2k_3.\xi_2\Tr(P_{-}\fsH_{(n)}M_p \gamma.k_2\gamma.\xi_{1})\bigg\}L_2,
\nonumber\\
{\cal A}_{3}&\sim&-2\bigg\{2t(k_3.\xi_2)\Tr(P_{-}\fsH_{(n)}M_p \gamma.k_3\gamma.\xi_{1})+2(-u-\frac{1}{4})k_1.\xi_2\Tr(P_{-}\fsH_{(n)}M_p \gamma.k_3\gamma.\xi_{1})\bigg\}L_3,
\nonumber\\
\labell{48}\eeqa

where

\beqa
L_1&=&(2)^{-2(t+s+u)}\pi{\frac{\Gamma(-u+\frac{1}{4})
\Gamma(-s+\frac{1}{4})\Gamma(-t+\frac{1}{2})\Gamma(-t-s-u+\frac{1}{2})}
{\Gamma(-u-t+\frac{3}{4})\Gamma(-t-s+\frac{3}{4})\Gamma(-s-u+\frac{1}{2})}},\nonumber\\
L_2&=&(2)^{-2(t+s+u)-1}\pi{\frac{\Gamma(-u-\frac{1}{4})
\Gamma(-s+\frac{3}{4})\Gamma(-t+1)\Gamma(-t-s-u)}
{\Gamma(-u-t+\frac{3}{4})\Gamma(-t-s+\frac{3}{4})\Gamma(-s-u+\frac{1}{2})}},\nonumber\\
L_3&=&(2)^{-2(t+s+u)-1}\pi{\frac{\Gamma(-u-\frac{1}{4})
\Gamma(-s+\frac{3}{4})\Gamma(-t)\Gamma(-t-s-u)}
{\Gamma(-u-t+\frac{3}{4})\Gamma(-t-s+\frac{3}{4})\Gamma(-s-u+\frac{1}{2})}}.
\nonumber\eeqa

The amplitude now satisfies Ward identity associated to the gauge field, which means that by replacing   $\xi_{2a}\rightarrow k_{2a}$  it gives  zero result.
One can write the amplitude by  doing more simplifications. Using momentum conservation along the world volume of brane  and making use of the various identities we obtain 
\beqa {\cal A}^{C\phi A T}&=&{\cal A}_{1}+{\cal A}_{2}+{\cal A}_{3},\labell{17u}\eeqa
where
\beqa
{\cal A}_{1}&\!\!\!\sim\!\!\!&2\xi_{1i}\xi_{2a}k_{3c}k_{2d}
\Tr(P_{-}\fsH_{(n)}M_p\Gamma^{cadi}
)L_1,
\nonumber\\
{\cal A}_{2}&\sim& \bigg\{-\Tr(P_{-}\fsH_{(n)}M_p \gamma.\xi_2\gamma.\xi_{1})(u+\frac{1}{4})-2k_3.\xi_2\Tr(P_{-}\fsH_{(n)}M_p \gamma.k_2\gamma.\xi_{1})\bigg\}L_3(2t)
\nonumber\\&&
+\Tr(P_{-}\fsH_{(n)}M_p \gamma.k_3\gamma.\xi_{1})\bigg\{2t(k_3.\xi_2)+2(-u-\frac{1}{4})k_1.\xi_2\bigg\}(-2L_3).
\labell{487}\eeqa

  $\fsH_{(n)}, M_p, \Gamma^{cadi}$ are constructed out of the antisymmetric
gamma matrices thus the amplitude is non zero for $p=n+1$ and $ p+1=n$ cases.
If we apply momentum conservation, we find

\beqa
s+t+u=-p^ap_a-\frac{1}{4},
\nonumber\eeqa

Regarding the facts that for non-BPS branes $p^ap_a \rightarrow \frac{1}{4}$ and the vertex of two scalars and one gauge field is non zero, we understand $t \rightarrow 0$.  Applying the above constraints,  we obtain the universal tachyon expansion. Indeed the following unique expansions should be held for all four point tachyon amplitudes:
\beqa
 t \rightarrow 0,  \quad  s \rightarrow  -\frac{1}{4} , \quad u \rightarrow  -\frac{1}{4}.
\labell{esio}\eeqa
as they have been argued in \cite{Hatefi:2012wj}. Notice that in terms of momenta the correct expansion is :
\beqa
  (k_1+k_2)^2 \rightarrow 0,  \quad  k_1.k_3 \rightarrow  0 , \quad k_2.k_3 \rightarrow  0,
\nonumber\eeqa

Now if we apply the expansion \reef{esio} to the all Gamma functions of the amplitude then  we reveal that  $C \phi A T$ amplitude has to have infinite massless scalar and  tachyon poles.  Before proceeding more several remarks have to be made.

 \vskip.1in

 It is argued that all four point functions including one RR, tachyon, gauge and scalar vertex operators must have infinite $t,s',u',(t+s'+u')$ channel poles, where  $s'=s+\frac{1}{4},u'=u+\frac{1}{4}$.
\vskip.1in

The important point for the $C \phi A T$ amplitude is that it is an exceptional amplitude. Due to the following reasons neither does it have  $s', (t+s'+u')$  tachyon, scalar nor  infinite poles. Due to the kinematic reasons, neither there is a coupling between two tachyons, one gauge and one scalar nor a coupling between two tachyons and one scalar field (even these amplitudes have non-zero CP factors). In fact conformal field theory techniques tell us that the amplitude  of $  TT\phi$ is vanished. Thus there is no $s'$ channel pole for $C\phi A T$ amplitude.  

 \vskip.1in

 By applying selections rules (based on string CP matrices), one understands that the amplitudes
  \beqa
  <V_{A^{0}} V_{T^{0}} V_{\phi^{-1}} V_{\phi^{-1}}>, <V_{A^{0}} V_{T^{0}} V_{A^{-1}} V_{\phi^{-1}}>.
  \nonumber\eeqa
  do not make sense in type II superstring theory because $\Tr(I\sig_1\sig_3\sig_3)=0$.
  \vskip.1in
  Therefore neither there are  single poles nor infinite poles of $s', (s'+t+u')$, as  the amplitude  \reef{48} does not carry the coefficients of $\Gamma(-s-\frac{1}{4}), \Gamma(-t-s-u-1/2)$ in its final form. Let us now analyse all infinite $u'$ tachyon and massless $t$  channel scalar poles of the amplitude.

  \vskip.2in

\section{Infinite tachyon poles of the $C\phi A T$ amplitude for $n=p+1$ case }

Given the explained universal expansion for tachyons, one reveals that the first and second terms of  \reef{48} do not contribute to any poles and in fact they are just contact  terms in which we are not interested in producing them (the method of producing all contact interactions in string theory has been explained in \cite{Hatefi:2012wj}).

\vskip .2in

On the other hand the third and fourth terms of \reef{48} have to be added to give rise to all infinite $u'$ channel tachyon poles and essentially the last term of  \reef{48} has infinite t-  channel scalar poles.
\vskip.1in

Let us write down all the infinite tachyon poles as below
\beqa
 4k_3.\xi_2  (k_{2a}+k_{3a}) \xi_{1i}\bigg(\Tr(P_{-}\fsH_{(n)}M_p \Gamma^{ai})\bigg) L_2,
\labell{411}\eeqa
One has to use momentum conservation along the brane's world volume. There is also an identity for RR as
\beqa
p_a\epsilon^{a_0...a_{p-1}a}=0.
\nonumber\eeqa
In order to be able to obtain all the infinite $u'$-channel tachyon poles  one has to apply all the identities and we find all tachyon poles as follows   
 \beqa
 -4k_3.\xi_2   k_{1a}  \xi_{1i}\bigg(\Tr(P_{-}\fsH_{(n)}M_p \Gamma^{ai})\bigg) L_2,
\labell{411}\eeqa
where the expansion of $L_2$ is
\beqa
L_2&=&-\pi^{3/2}\bigg(\frac{1}{u'}\sum_{n=-1}^{\infty}c_n(s'+t)^{n+1}+\sum_{p, n,m=0}^{\infty}h_{p,n,m}(u')^{p}(t s')^{n}(t+s')^m\bigg),\labell{high66}\eeqa
some of the coefficients are
\beqa
&&c_{-1}=1,\,c_0=0,\,c_1=\frac{1}{6}\pi^2,h_{0,0,1}=\frac{1}{3}\pi^2,h_{1,0,1}=h_{0,0,2}=6\z(3).
\nonumber
\eeqa
\vskip.1in



In this section by making use of the  infinite corrections of one RR, one tachyon and one scalar field \reef{highaa},  we want to produce all the infinite tachyon poles and to actually show that those obtained corrections are exact and have no on-shell ambiguity. Let us write down infinite tachyon $u'$-channel poles (there is no $s'$ channel pole for this amplitude) as

\beqa
&&-(4k_3.\xi_2)k_{1a}\xi_{1i}\frac{16}{(p)!}(\pi^{3/2})(\mu'_p\beta'\pi^{1/2})\sum_{n=-1}^{\infty}c_n\frac{1}{u'}(s'+t)^{n+1}\nonumber\\&&\times
 H^{i}_{a_{0}\cdots a_{p-1}}
\eps^{a_{0}\cdots a_{p-1}a }\Tr(\lam_1\lam_2\lam_3).\label{bbx}\eeqa

The above amplitude is  normalised by a coefficient of $(\mu'_p\beta'\pi^{1/2})$.  The following field theory amplitude should be considered
\beqa
{\cal A}&=&V^{\alpha}(C_{p},\phi_1,T)G^{\alpha\beta}(T)V^{\beta}(T,T_3,A_2).\labell{amp42}\eeqa

We also need to take into account several remarks. 

\vskip.1in

Tachyon propagator ($G^{\alpha\beta}(T)$) is derived from  the kinetic term of  the tachyon. To obtain  $V^{\beta}(T,T_3,A_2)$ one has to  extract  the covariant derivative of tachyon ($D_aT=\partial_T-i[A^a,T]$) inside the non-Abelian kinetic term of tachyon. $V^{\alpha}(C_{p},\phi_1,T)$ is derived  from the coupling  \reef{jj}, such that

\beqa
V^{\beta}(T,T_3,A_2)&=&iT_p(2\pi\alpha')(k_3-k).\xi_2\Tr(\lam_2\lam_3\Lambda^\beta)=2iT_p(2\pi\alpha')k_3.\xi_2
\Tr(\lam_2\lam_3\Lambda^\beta),\nonumber\\
V^{\alpha}(C_{p},\phi_1,T)&=&2\mu'_p\beta'\frac{(2\pi\alpha')^{2}}{(p)!}\epsilon^{a_0\cdots a_{p}}H^{i}_{a_0\cdots a_{p-1}}k_{a_p}\xi_{1i}\Tr(\lam_1\Lambda^{\alpha}),\nonumber\\
G^{\alpha\beta}(T) &=&\frac{-i\delta^{\alpha\beta}}{(2\pi\alpha') T_p
(k^2+m^2)}=\frac{-i\delta^{\alpha\beta}}{(2\pi\alpha') T_p
(u')}.
\label{moi}\eeqa

In the vertex of $V^{\beta}(T,T_3,A_2)$, we have used the momentum conservation on the right hand side of the Feynman rule $(k_3+k_2+k)^a=0$,  in which $k$ is the momentum of tachyon propagator. If we replace the above vertices \reef{moi} inside the field theory amplitude \reef{amp42}, use the momentum conservation in left hand side of the Feynman rule, namely $((k_1+p)^a=k^a$) and take into account the first coefficient of the expansion $c_{-1}=1$  in \reef{bbx} then  we are able to exactly produce  the first simple tachyon $u'$-channel pole; however, as it is seen from  \reef{bbx}, the amplitude involves infinite $u'$ channel tachyon poles. In order to deal with those infinite poles, one has to use some of  the arguments that appeared in \cite{Hatefi:2013mwa}. Basically both of the simple tachyon pole and $V^{\beta}(T,T_3,A_2)$ do not receive any correction.
Because the kinetic term of tachyon  has been fixed in tachyon DBI action and therefore it  has no correction. Hence in order to produce those infinite tachyon poles, one has to find out the infinite corrections to one RR (p-form), one tachyon and one on-shell scalar. In fact these corrections  have been derived by performing the scattering amplitude of $C\phi T$ in \reef{highaa}.

\vskip.1in

Once we apply all the infinite higher derivative corrections of  $\partial_i C_{p} \wedge DT \phi^i $ \reef{highaa}, we can find

\beqa
V^{\alpha}(C_{p},T,\phi_1)=\frac{2\mu'_p\beta'(2\pi\alpha')^{2} k_{a_p}\xi_{1i}}{(p)!}\epsilon^{a_0\cdots a_{p}}H^{i}_{a_0\cdots a_{p-1}}\sum_{m=-1}^{\infty}c_m(\alpha'k_1\cdot k)^{m+1}\Tr(\lam_1\Lambda^{\alpha}).\label{2300}\eeqa

Now if we use the fact that $\sum_{m=-1}^{\infty}c_m(\alpha'k_1\cdot k)^{m+1}=\sum_{m=-1}^{\infty}c_m (t+s+1/4)$ and  substitute \reef{2300} inside the field theory amplitude \reef{amp42} (while keep the other vertices fixed) then we are exactly able to explore all the infinite $u'$ channel poles of \reef{bbx}.
\vskip.2in 

This  clearly  shows that the obtained higher derivative corrections of \reef{highaa} are exact and have no on-shell ambiguity. Note also that there are no residual contact interactions left over in producing an infinite number of the tachyon poles of the $C\phi A T$ S-Matrix.

\section{Infinite scalar t-channel poles of the $C\phi A T$ amplitude for $n=p+1$ case }

The last term of  \reef{48} includes an infinite number of  t-channel massless scalar poles as follows
\beqa
 -4k_1.\xi_2  k_{3a} \xi_{1i}\bigg(\Tr(P_{-}\fsH_{(n)}M_p \Gamma^{ai})\bigg) (-u-1/4)L_3,
\labell{413}\eeqa
 and the expansion of  $(-u-1/4)L_3$  is

 \beqa
(-u-1/4)L_3&=&-\pi^{3/2}\bigg(\frac{1}{t}\sum_{n=-1}^{\infty}c_n(u'+s')^{n+1}
+\sum_{p,n,m=0}^{\infty}h_{p,n,m}t^{p}(u' s')^{n}(u'+s')^m\bigg),\label{tpoles}\eeqa
where  $c_n$ coefficients are precisely the coefficients that have been written down in the momentum expansion of the S-matrix element of $C\phi T$.  Some other coefficients of $c_n, h_{p,n,m}$ can be summarised as
\beqa
&& c_2=2\z(3), h_{2,0,0}=h_{0,1,0}=2\z(3),h_{1,0,0}=\frac{1}{6}\pi^2,h_{1,0,2}=\frac{19}{60}\pi^4,\nonumber\\
&&h_{0,0,1}=\frac{1}{3}\pi^2,h_{0,0,3}=e_{2,0,1}=\frac{19}{90}\pi^4,h_{1,1,0}=h_{0,1,1}=\frac{1}{30}\pi^4.\labell{865} \nonumber\eeqa

\vskip.1in


In  order to follow this section two main goals  are needed.

\vskip.1in

 1) We would like to explicitly show that the  infinite corrections of one RR, one tachyon and one scalar field that  obtained in \reef{highaa} are exact and do work.

 2) We are going to show that there are infinite t-channel scalar poles which must be obtained in field theory by applying the correct higher derivative corrections. In particular we want to show that our expansion in \reef{esio} is unique because by using it we can precisely produce an infinite number of the scalar poles of the $C \phi A T$ string amplitude. If we insert the first term of the expansion of $(-u-1/4)L_3$ into the amplitude then  we can just write down the infinite  $t$-channel scalar poles  of the amplitude (there is no ($s'+t+u'$) scalar pole for this amplitude) as below
\beqa
&& -4k_1.\xi_2  k_{3a} \xi_{1i} \frac{16}{(p)!}(\pi^{3/2})(\mu'_p\beta'\pi^{1/2})\sum_{n=-1}^{\infty}c_n\frac{1}{t}(s'+u')^{n+1}\nonumber\\&&\times
 H^{i}_{a_{0}\cdots a_{p-1}}
\eps^{a_{0}\cdots a_{p-1}a }\Tr(\lam_1\lam_2\lam_3).\label{bbx3}\eeqa
The rule for  the  field theory amplitude for this case is
\beqa
{\cal A}&=&V^{\alpha}(C_{p},T_3,\phi)G^{\alpha\beta}(\phi)V^{\beta}(\phi,\phi_1,A_2),\labell{amp4666}\eeqa
Now we are going to make some remarks about field theory analysis. Scalar propagator ($G^{\alpha\beta}(\phi)$) should be found from  the kinetic term of the scalar.  $V^{\beta}(\phi,\phi_1,A_2)$ must be derived by extracting the covariant derivative of the scalar field  in the non-Abelian kinetic term of the scalars ($D_a \phi^i=\partial_a \phi^i-i[A^a,\phi^i]$).  $V^{\alpha}(C_{p},T,\phi)$ is related to the previous S-matrix , \ie the coupling \reef{jj} and  $V^{\beta}(\phi,\phi_1,A_2)$ is derived in \cite{Hatefi:2012ve}. Let us write them down in below :

\beqa
V^{\beta}(\phi,\phi_1,A_2)&=&-2i (2\pi\alpha')^2 T_p k_1.\xi_2 \xi_{1j} \Tr(\lam_1\lam_2\Lambda^{\beta}),\label{mmmpo}\eeqa
\beqa
V^{\alpha}(C_{p},\phi,T_3)&=&2\mu'_p\beta'\frac{(2\pi\alpha')^{2}}{(p)!}\epsilon^{a_0\cdots a_{p}}H^{i}_{a_0\cdots a_{p-1}}k_{3_a{p}}\Tr(\lam_3\Lambda^{\alpha}),\nonumber\\
G^{\alpha\beta}(\phi) &=&\frac{-i\delta^{\alpha\beta}\delta^{ij}}{(2\pi\alpha')^2 T_p
(k^2)}=\frac{-i\delta^{\alpha\beta}\delta^{ij}}{(2\pi\alpha')^2 T_p
(t)}.
\label{moi3}\eeqa

\vskip.1in

 $k$ is the momentum of off-shell scalar propagator. If we replace the  vertices \reef{mmmpo} and   \reef{moi3} inside the field theory amplitude \reef{amp4666}, use the momentum conservation for the left and right hand sides of Feynman rule  and consider $c_{-1}=1$  in \reef{bbx3} then  we are precisely  able to find out  the first simple massless scalar $t$-channel pole of the string amplitude of \reef{bbx3} in field theory side; however, as it is seen from $(-u-1/4)L_3$ expansion \reef{tpoles}, the amplitude includes an infinite number of t-channel scalar poles. 
 
 In order to produce those singularities in field theory amplitude, one has to apply several points. Basically both simple t-channel scalar pole and $V^{\beta}(\phi,\phi_1,A_2)$ do not receive any corrections because  the kinetic term of scalars does not obtain any correction  as it has been fixed in  standard DBI action and  also simple pole does not receive any correction.  Thus in order to look for those infinite massless scalar poles one needs to look for infinite corrections to one RR (p-form), one on-shell tachyon and one off-shell scalar field.  Surprisingly these corrections  can be derived from  the  scattering amplitude of $C\phi T$ in \reef{highaa}.
\vskip.1in

Hence, if we use all the infinite  higher derivative corrections of  $\partial_i C_{p} \wedge DT \phi^i $ \reef{highaa} then we can explore all order extensions of the $V^{\alpha}(C_{p},\phi,T_3)$ vertex operator as follows
\beqa
V^{\alpha}(C_{p},\phi,T_3)&=&2\mu'_p\beta'\frac{(2\pi\alpha')^{2}}{(p)!}\epsilon^{a_0\cdots a_{p}}H^{i}_{a_0\cdots a_{p-1}} k_{3_{a_p}}\sum_{m=-1}^{\infty}c_m(\alpha'k_3\cdot k)^{m+1}\Tr(\lam_3\Lambda^{\alpha}),\label{2306}\eeqa

\vskip.1in

We use momentum conservation in world volume direction such that  $\sum_{m=-1}^{\infty}c_m(\alpha'k_3\cdot k)^{m+1}=\sum_{m=-1}^{\infty}c_m (u+s+1/2)$, substitute \reef{2306} inside the field theory amplitude \reef{amp4666} and simultaneously keep the form of the other vertices fixed. If we do so then we are able to find out :

\beqa
&& -4k_1.\xi_2  k_{3a_{p}} \xi_{1i} \frac{16}{(p)!}\pi^2\mu'_p\beta'\sum_{n=-1}^{\infty}c_n\frac{1}{t}(s+u+1/2)^{n+1}\nonumber\\&&\times
 H^{i}_{a_{0}\cdots a_{p-1}}
\eps^{a_{0}\cdots a_{p}}\Tr(\lam_1\lam_2\lam_3).\label{bbx67}\eeqa

which is exactly all the infinite t-channel massless scalar poles of \reef{bbx3}.  Hence this again shows that the derived higher derivative corrections of \reef{highaa} are exact and  have no on-shell ambiguities. Notice that in
exploring an infinite number of t-channel scalar  poles of  the $C\phi A T$ S-Matrix there are no residual contact interactions left over. This can be understood by comparing field theory amplitude \reef{bbx67} with string theory S-matrix elements \reef{bbx3}.
 \vskip.1in
Finally note that the method of extracting all the contact terms in string theory has been extensively pointed out in \cite{Hatefi:2012wj,Hatefi:2012rx}; however, it is worth pointing out the following remarks.
\vskip.1in
In order to look for an infinite number of contact interactions of the $C \phi A T$ (for $p+1=n$ case) the following coupling has to be considered 
\beqa
\int_{\Sigma_{p+1}} \partial^i C_p \wedge DT \phi^i
\nonumber\eeqa

 particularly one has to extract the covariant derivative of tachyon and  take into account the commutator inside the covariant derivative of tachyon $([A^a,T])$. More importantly one should add together all the contact interactions coming from the second parts of  $L_2$ and $(-u') L_3$ expansions in (31) and (37) accordingly.

 \section{Conclusions}

In this paper based on internal CP matrix of the strings in the presence of non-BPS branes, we have constructed the  selection rules for all non-BPS amplitudes.
We observed that,  these selection rules  have the significant potential to rule out several non-BPS higher point correlation functions of type IIA (IIB) string theory without the need for taking integrals of the world-sheet. More significantly these selection rules showed us that there should not be any couplings for some of the non-BPS open/closed strings in their effective actions. Based on the suggested rules we realise that what kinds of couplings are not allowed in these theories. In section three by explicit calculations  we discovered all order $\alpha'$ corrections of one RR, one tachyon and one scalar field in the world volume of non-BPS branes.
 \vskip .1in

In particular we obtained a unique expansion for tachyon amplitudes. This expansion is very useful because by applying it to the string amplitudes, one  can find out all singularities of non-BPS  higher point functions of the string theory  without the need for knowing the  complete results of the world-sheet integrals.
This idea clearly has been applied to the amplitude of one RR and four tachyons in the world volume of $D\bar D$ system but has not been publicized yet
\cite{Hatefi:2013eh}.
\vskip .1in

In section four using selection rules and universal tachyon expansion we dealt with the amplitude of one RR, one scalar, one gauge field and one tachyon in the world volume of non-BPS branes. By applying selection rules to the field theory amplitude we revealed that  neither  does the $C \phi A T$ amplitude have single /double/ infinite $(s')$ tachyon nor $(s'+u'+t)$ poles. Thus it is certainly an exceptional S-matrix.
\vskip .1in
Essentially  by making use of the selection rules, the  universal tachyon expansion and the derived  corrections of one RR, one tachyon and one scalar field (results in section four), we were able to produce an infinite number of $u'$ tachyon and $t$-channel massless scalar field poles of the $C \phi A T$ amplitude and their higher derivative corrections. It is of high importance to mention that the Wess-Zumino effective actions of the unstable branes give  all order corrections of the related string amplitudes.

\section*{Acknowledgments}

The author would like to thank J.Polchinski, A. Sen , R.Myers, K.S.Narain, N.Lambert, F.Quevedo, N.Arkani-Hamed, G.Veneziano, G.Moore, A.Sagnotti  and L.Alvarez-Gaume for  valuable comments. He also thanks E.Martinec and P.Horava for very useful discussions during recent CMT workshop at ICTP.

\end{document}